\documentclass[pdftex,twocolumn,epjc3]{svjour3}

\usepackage[utf8]{inputenc}
\usepackage{amsmath}
\usepackage{amssymb}
\usepackage{graphicx}
\usepackage[colorlinks,citecolor=blue,urlcolor=blue,linkcolor=blue]{hyperref}
\usepackage[numbers,sort&compress]{natbib} 
\usepackage[export]{adjustbox}
\usepackage[usenames,dvipsnames,svgnames,table]{xcolor}
\usepackage{multirow}

\newcommand{\fref}[1]{Fig. \ref{#1}}

\newcommand{\tref}[1]{Tab.~\ref{#1}}

\allowdisplaybreaks 

\journalname{Eur. Phys. J. C}

\begin{document} 

\title{Apparent convergence in functional glueball calculations}

\author{Markus Q.~Huber\thanksref{e1,addr1} \and Christian S. Fischer\thanksref{e2,addr1,addr2} \and H\`elios Sanchis-Alepuz\thanksref{e3,addr3}}

\thankstext{e1}{e-mail: markus.huber@theo.physik.uni-giessen.de}
\thankstext{e2}{e-mail: christian.fischer@theo.physik.uni-giessen.de}
\thankstext{e3}{e-mail: helios.sanchis-alepuz@silicon-austria.com}

\institute{Institut f\"ur Theoretische Physik, Justus-Liebig-Universit\"at Giessen, Heinrich-Buff-Ring 16, 35392 Giessen, Germany\label{addr1}
          \and
          Helmholtz Forschungsakademie Hessen f\"ur FAIR (HFHF), GSI Helmholtzzentrum f\"ur Schwerionenforschung, Campus Gie{\ss}en, 35392 Gie{\ss}en, Germany\label{addr2}
          \and
          Silicon Austria Labs GmbH, Sandgasse 34, 8010 Graz, Austria\label{addr3}
}

\date{\today}

\maketitle

\begin{abstract}
We scrutinize the determination of glueball masses in pure Yang-Mills theory from functional equations, i.e. Dyson-Schwinger
and Bethe-Salpeter equations. We survey the state-of-the-art input (dressed propagators and vertices) with an emphasis on the
stability of the results under extensions of the employed truncations and explore the importance of different aspects 
of the bound state equations, focusing on the three lightest glueballs with $J^\textsf{PC}=0^{++}$, $0^{-+}$ and $2^{++}$.
As an important systematic extension compared to previous calculations we include two-loop diagrams in the Bethe-Salpeter 
kernels. In terms of the glueball
spectrum we find only marginal mass shifts compared to previous results, indicating apparent convergence of the system. 
As a by-product, we also explore gauge invariance within a class of Landau-type gauges. 

\end{abstract}

\section{Introduction}

The calculation of the spectrum of hadrons is one of the main tests of quantum chromodynamics (QCD) as the theory of the strong interaction, but it depends naturally also on our grasp and mastery of the employed methods.
While different approaches are successful for certain hadrons and lead to coherent results, some sectors still remain obscure.
Among them are glueballs \cite{Fritzsch:1972jv,Fritzsch:1975tx} which are both experimentally and theoretically elusive \cite{Klempt:2007cp,Crede:2008vw,Mathieu:2008me,Ochs:2013gi,Llanes-Estrada:2021evz,Vadacchino:2023vnc}.
One of the main challenges is the mixing (for certain quantum numbers) with quark-antiquark states.
This can be avoided in theoretical calculations by making quarks infinitely heavy and considering pure gauge theory.
For this scenario, lattice methods have established a spectrum of purely gluonic states \cite{Bali:1993fb,Morningstar:1999rf,Chen:2005mg,Athenodorou:2020ani} which serve as convenient benchmark for other methods.
Despite neglecting the effects of quarks, such calculations not only provide useful guidance for the understanding of glueballs 
in full QCD but also a good first qualitative or even semi-quantitative estimate for real glueballs \cite{Gregory:2012hu,Morningstar:2024vjk}.
Experimental evidence for the existence of such states has been claimed in recent analyses of BESIII data in the case 
of scalar \cite{Sarantsev:2021ein,Rodas:2021tyb} and pseudoscalar glueballs \cite{BESIII:2023wfi}.
The situation for the tensor glueball is more complicated \cite{Klempt:2022qjf}.

Functional methods, which are successfully used to calculate spectra of baryons and mesons, see, e.g., \cite{Cloet:2013jya,Eichmann:2016yit}, were only recently able to catch up with lattice methods and 
provide a concise spectrum of pure glueballs with quantum numbers $J=0,2,3,4$, $\textsf{P}=\pm1$ and 
$\textsf{C}=1$ \cite{Huber:2020ngt,Huber:2021yfy}. A crucial element in these calculations has been 
high quality input \cite{Huber:2020keu} in terms of propagators and vertices from the underlying Yang-Mills 
theory.\footnote{Earlier attempts relied on models and their predictive power was limited due to the 
severe model parameter dependence \cite{Meyers:2012ka,Sanchis-Alepuz:2015hma,Souza:2019ylx,Kaptari:2020qlt}.}

\begin{figure*}[tb]
	\begin{center}
	\includegraphics[width=0.9\textwidth]{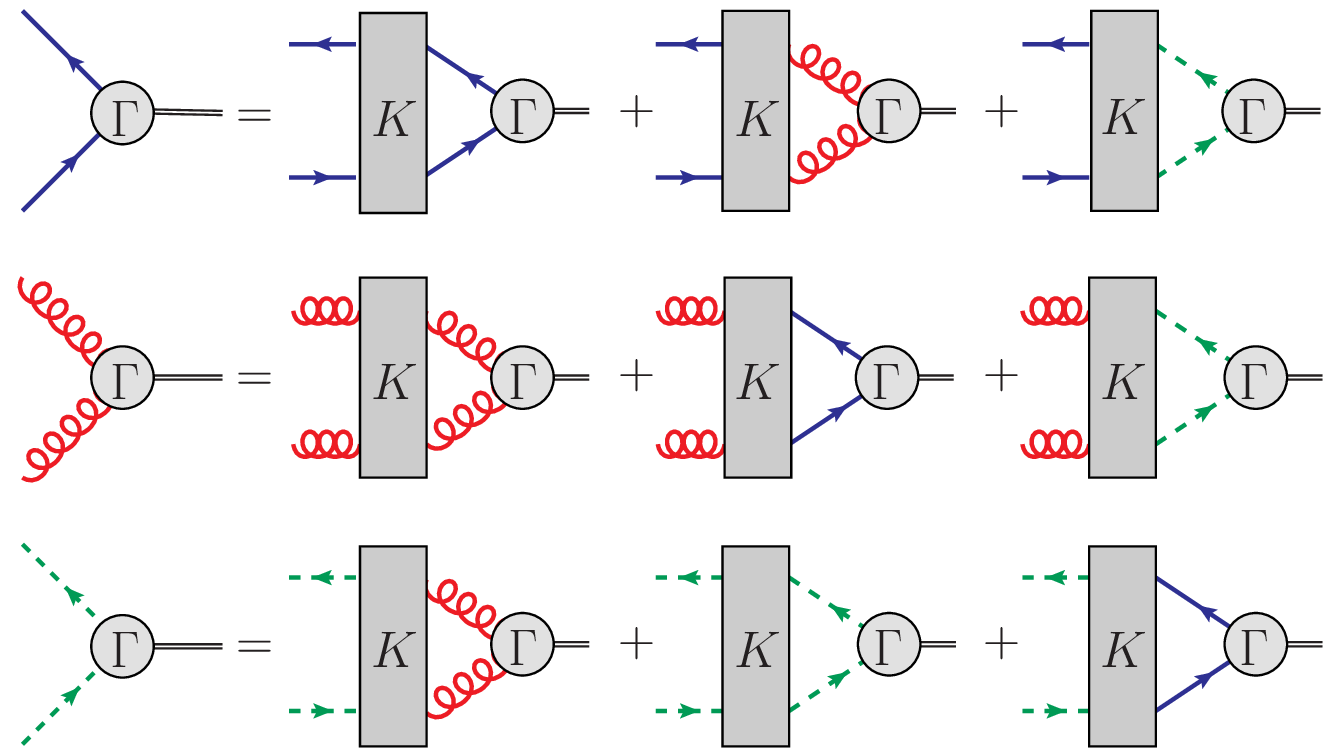}
	\caption{
		The coupled set of BSEs for a glueball made from two gluons and a pair of Faddeev-Popov (anti-)ghosts and for a conventional meson made of a quark-antiquark pair.
		Wiggly lines denote dressed gluon propagators, dashed lines dressed ghost propagators and solid lines dressed quark propagators.
		The gray boxes represent interaction kernels. The Bethe-Salpeter amplitudes $\Gamma$ are denoted by gray disks.
	}
	\label{fig:bses}
	\end{center}
\end{figure*}
The underlying truncation scheme 
is based on the 3PI effective action
truncated at three loops. In this work, we review variations of this setup, most notably the effect of extensions, 
and show that the results exhibit a remarkable stability under these changes.
The kernels of the bound state equations can be derived in the same truncation and lead to one-particle exchange diagrams and one-loop diagrams.
The latter, however, were not included in Refs.~\cite{Huber:2020ngt,Huber:2021yfy} due to computational complexity.
In this work, we remedy that and present the results including two-loop diagrams in the kernels for the three lightest glueballs.
Since the calculations are very expensive in terms of CPU time, we used these as proxy to address potential quantitative changes 
induced by the two-loop diagrams, but refrained at this stage from calculating the whole spectrum.
We also discuss the hierarchy of diagrams and the effect of kinematic approximations for the three-gluon vertex to identify computationally simpler setups.
Finally, we test the dependence of these glueball masses on variations of the input in form of available solutions that differ only in the low momentum regime.

We discuss the underlying equations in Sec.~\ref{sec:3PI}.
In Sec.~\ref{sec:correlationFunctions}, we present the input and discuss its stability under various modifications.
The main result is the fully consistent calculation of the bound states in Sec.~\ref{sec:spectrum} including various tests.
We close with a summary in Sec.~\ref{sec:summary}.
The appendix contains details on the scale setting.

\section{The 3PI effective action}
\label{sec:3PI}

\begin{figure*}[tb]
	\begin{center}
	\includegraphics[width=0.98\textwidth]{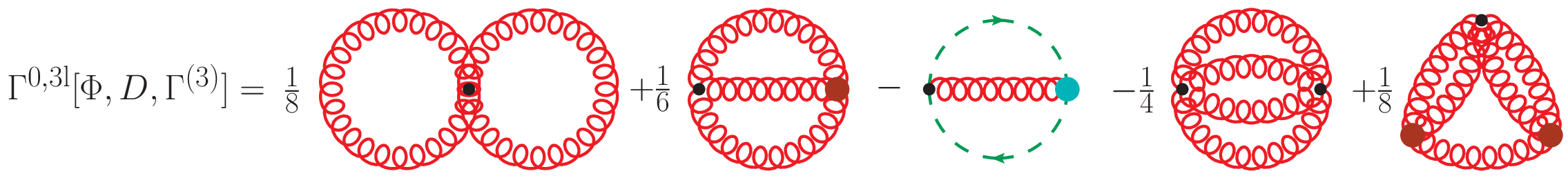}\\
	\hskip-1.3cm\includegraphics[width=0.90\textwidth]{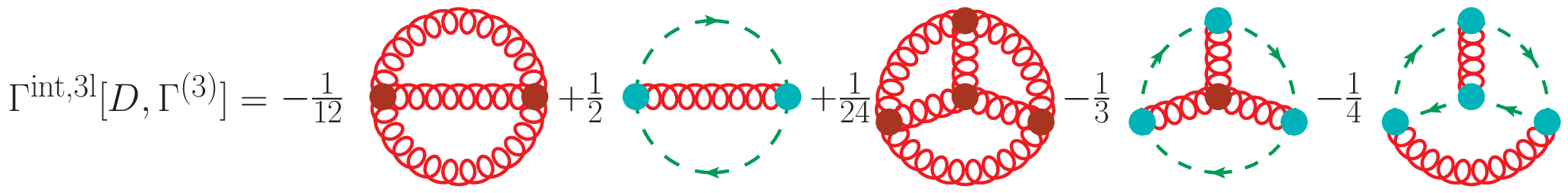}
	\caption{
		The 3PI effective action truncated to three loops.
		The meaning of lines is described in \fref{fig:bses}.
		Dots represent bare and disks dressed vertices.
		Shown is the Yang-Mills part only, but the quark contributions can be inferred from the ghost ones which have the same structure.
	}
	\label{fig:3PI_EA}
	\end{center}
\end{figure*}

In general, solving bound state Bethe-Salpeter equations (BSEs) requires input in terms of propagators and 
vertices. These can be determined 
from the underlying tower of functional equations of the theory as given by a set of Dyson-Schwinger equations (DSEs)
(or the functional renormalization group). In the quark sector of QCD, simple models for these correlation functions
like the well-studied rainbow-ladder approach provide reasonable results for a wide range of observables like 
spectra or form factors, see e.g. 
\cite{Roberts:1994dr,Maris:1999nt,Qin:2011dd,Cloet:2013jya,Eichmann:2016yit} and references therein.

However, for glueballs the situation is different. The prime reason is the gluon self-interaction, which 
entails a much more complex structure of the DSE for the gluon propagator compared to the one of the quark.
Consequently, a useful truncation equivalent to rainbow-ladder does not exist. In addition, the interplay 
of the gluon with ghost fields appearing due to gauge fixing needs to be taken into account explicitly.
As a consequence, the glueball BSE (even without quarks) has a complex structure with more than one interaction
diagram, see \fref{fig:bses} for details. In these equations, the three-gluon vertex plays a very prominent role
as will be discussed explicitly in Sec.~\ref{sec:spectrum}.

Functional glueball calculations for $J=0$ using models for propagators and/or vertices were carried out in the 
past, but they were not able to provide a consistent picture for scalar and pseudoscalar glueballs
\cite{Meyers:2012ka,Sanchis-Alepuz:2015hma}. In contrast, high quality results in agreement with lattice Yang-Mills 
theory have been found once a consistent truncation for the underlying DSEs and the bound state equations
has been employed \cite{Huber:2020ngt} with propagators and vertices determined self-consistently in this truncation. 
This type of consistency seems to be a decisive and common feature of all high quality bound state calculations.

The general bound state equation for glueballs is shown in \fref{fig:bses}. To highlight the mixing with quarks 
in full QCD, the corresponding diagrams are also shown, but we will focus on pure Yang-Mills theory from here on.
Then, all diagrams containing quarks can be dropped and a system of two equations with two diagrams each remains.
To distinguish the two equations, we will refer to them as the gluon and ghost BSEs, respectively, but it should 
be noted that they jointly describe a glueball.\footnote{In all our calculations we use Landau gauge, which is 
the best explored gauge with functional and lattice methods, see \cite{Huber:2018ned} for more details.
The relaxation to general linear covariant gauges is surely of interest to test the gauge parameter independence of the results, but current results, see, e.g.,~\cite{Aguilar:2015nqa,Capri:2015pja,Huber:2015ria,Aguilar:2016ock,Napetschnig:2021ria,Silva:2018ta,Cucchieri:2018doy}, need to be improved to reach the same level of sophistication.}

The truncation scheme we consider here is based on a nonperturbative loop expansion of the 3PI effective action with
dressed propagators and vertices (sometimes also called a skeleton expansion). Specifically, we keep terms up to three loops.
Such an expansion is much more systematic than typical 1PI truncations which act at the level of correlation functions.
The 3PI expansion offers the distinct possibility to perform self-consistent calculations on a given loop-level and provides for 
a systematic road to go to the next 'order' of expansion. Since it is not a perturbative expansion with a small parameter, 
there is no guarantee that contributions of the next order are smaller than contributions of the previous one. However, in 
practice it appears as if convergence has been reached on the three-loop level. Hence the term 'apparent convergence' in the title of this work. We will come back to this point below.

\begin{figure*}[tb]
	\includegraphics[height=1.6cm]{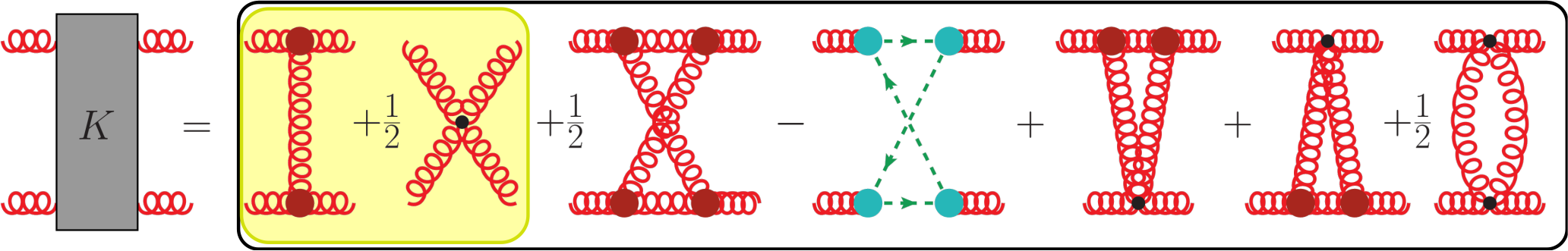}\\
	\vskip2mm
	\includegraphics[height=1.6cm]{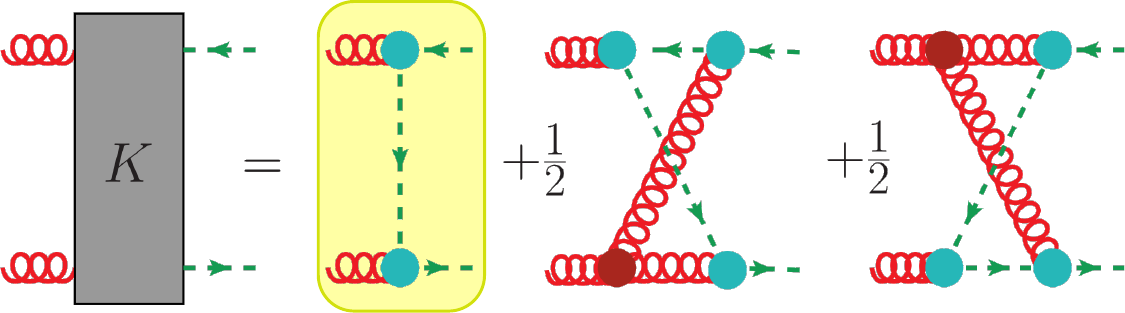}\hfill
	\includegraphics[height=1.6cm]{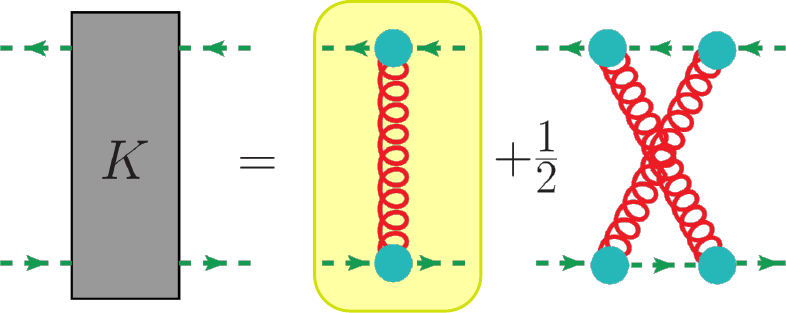}\hfill
	\includegraphics[height=1.6cm]{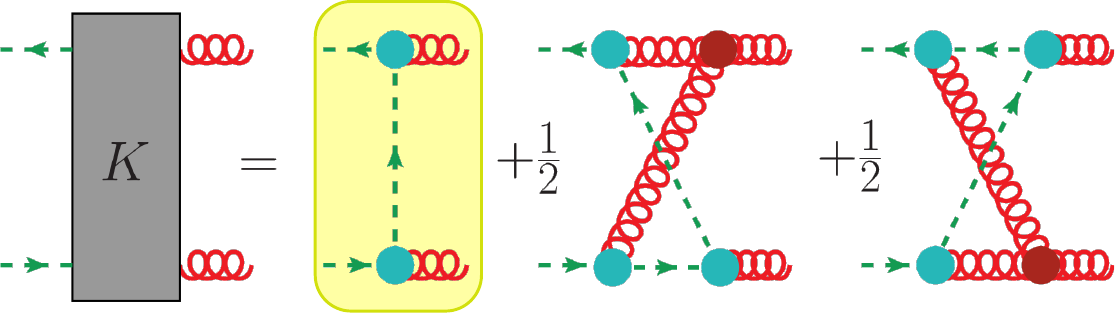}
	\caption{
		Interaction kernels from the 3PI effective action truncated at three loops.
		All propagators are dressed; disks represent dressed vertices, dots bare ones. Yellow boxes indicate the one-loop truncation, the black box the two-loop diagrams taken into account in this work.
		\label{fig:kernels}
	}
\end{figure*}

Conventionally, the 3PI effective action is split into two parts,
\begin{align}
	\Gamma[D,\Gamma^{(3)}]=\Gamma^{0}[D,\Gamma^{(3)}]+\Gamma^{\text{int}}[D,\Gamma^{(3)}],
\end{align}
which are shown in \fref{fig:3PI_EA} for the three-loop truncation \cite{Berges:2004pu,Carrington:2010qq}.
The 3PI effective action depends explicitly on the propagators and three-point functions, here denoted generically by $D$ and $\Gamma^{(3)}$, respectively.
The interaction kernels of the BSE $K$ can be derived from the effective action by taking two functional derivatives with respect to the propagators \cite{Fukuda:1987su,Komachiya:1989kc,McKay:1989rk}, see also \cite{Sanchis-Alepuz:2015tha} for a contemporary discussion and \cite{Huber:2020ngt} for the glueball case.
Such derivatives lead to many diagrams, but using the equations of motion of the three-point functions \cite{Berges:2004pu,Carrington:2010qq}, they can be resummed resulting in the diagrams shown in \fref{fig:kernels}.

There are two main differences to a rainbow-ladder type truncation .
First, one-loop diagrams appear in the 
interaction kernels which lead to two-loop bound state equations. Second, due to the aforementioned resummation, 
the one-particle exchange diagrams in the kernels have both vertices dressed in contrast to only one dressed 
vertex in a ladder type truncation. Initially, we only considered the one-particle exchange diagrams and the 
one with the four-gluon vertex in the kernels (marked by yellow boxes in \fref{fig:kernels})
\cite{Huber:2020ngt,Huber:2021yfy}. Preliminary results including additional diagrams (indicated by the black box) 
have already been reported in contributions for proceedings \cite{Huber:2021zqk,Huber:2023mls,Huber:2022rhh}.
Here, we elaborate on these calculations in Sec.~\ref{sec:spectrum}.

What also distinguishes our setup from a conceptional rainbow-ladder one with model input is the use of 
propagators and three-point functions \textit{calculated} from the 3PI effective action truncated at three-loops. 
The corresponding equations of motion (EOM) for these correlation functions correspond to stationarity conditions of the 3PI effective 
action with respect to the desired correlation function \cite{Berges:2004pu,Carrington:2010qq,Sanchis-Alepuz:2015tha}.
The resulting 3PI EOMs are very similar in structure to DSEs, which are equations of motion derived from the 1PI 
effective action. However, there are also crucial differences. While the DSEs each feature a finite number of diagrams 
but form an infinite tower of equations, the corresponding 3PI EOMs form a finite set but with (possibly) infinitely 
many diagrams due to the loop expansion. For want of a catchy name, the equations of motion from an $n$PI 
effective action are often referred to as DSEs for $n>1$ as well. This also seems justified as the equations of motion 
from different effective actions are identical in some cases. In the three-loop truncation used here, the propagator
equations indeed are equal for 1PI and 3PI except for the four-gluon vertex which can be dressed in the former 
and is bare in the latter. For three-point functions, the three-loop truncation of the 3PI effective action 
restricts the diagrams in the equations to one-loop, whereas the DSEs can have two-loop diagrams. The one-loop 
diagrams are topologically identical, but the internal vertices are all dressed in the 3PI case whereas in the 
1PI case there is always one bare vertex in each diagram. Further details are discussed in the review article 
\cite{Huber:2018ned}, which also offers a graphical representation of the corresponding equations. 
Explicit results for correlation functions in the 3PI setup are discussed in the next section.

\section{Correlation functions}
\label{sec:correlationFunctions}

As seen in \fref{fig:kernels}, the glueball bound state equations probe four different (nonperturbative) correlation functions:
the gluon and ghost propagators as well as the three-gluon and ghost-gluon vertices.
These are determined from the equations of motion of the 3PI effective action truncated at three loops, see the discussion above.
Corresponding results were reported in Ref.~\cite{Huber:2020keu}. Here, we elaborate on the stability 
of these results. To this end, we will compare them with results from other methods in Sec.~\ref{sec:comparison} 
and list various extensions and their impact on the results in Sec.~\ref{sec:extensions}.

\subsection{Comparison with other methods}
\label{sec:comparison}

The correlation functions of Yang-Mills theory have been investigated by a wide range of nonperturbative methods.
For conciseness we only consider self-contained functional calculations and discuss results from 1PI Dyson-Schwinger
equations \cite{Huber:2020keu}, the 3PI effective action \cite{Huber:2020keu} and the functional renormalization 
group\footnote{Investigations with respect to apparent convergence have also been performed within the FRG framework, see \cite{Ihssen:2024miv} and references therein.} (FRG) \cite{Cyrol:2016tym,Pawlowski:2022oyq}. A complementary approach is gauge fixed lattice gauge theory,
which provides benchmarks for various correlation functions. In the following, we will systematically 
compare results from each of these methods.

An additional complication in gauge fixed Yang-Mills theory is provided by the existence of a one-parameter family
of solutions for correlation functions, which are conveniently selected by choosing a value for 
the ghost dressing function at vanishing momentum, $G(0)$, as renormalization condition for the ghost propagator. 
Thus, we display bands of results in some figures which reflect this degree of freedom.
For more details on the solutions used here we refer to \cite{Huber:2020keu} and the general discussions 
of this topic in, e.g., \cite{Boucaud:2008ji,Fischer:2008uz,Maas:2009se,Alkofer:2008jy,Huber:2018ned}. In the context of 
this work, we discuss a possible interpretation of this family of solutions in Sec.~\ref{sec:family}.

\subsubsection{Propagators}

The propagators are the simplest correlation functions in terms of kinematics and tensor structures as they only 
depend on one momentum and have (in Landau gauge) only one dressing function each.\footnote{
For the gluon propagator, the appearance of quadratic divergences requires some special care \cite{Huber:2014tva,Huber:2018ned,Eichmann:2021zuv,Huber:2020keu} which makes solving the gluon equation more complicated than that of the ghost.}

\begin{figure}[tb]
	\includegraphics[width=0.48\textwidth]{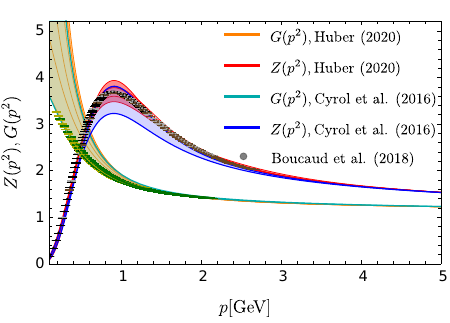}
	\caption{Gluon dressing function $Z(p^2)$ and ghost dressing function $G(p^2)$ from DSEs \cite{Huber:2020keu} (Huber (2020)) and FRG \cite{Cyrol:2016tym} (Cyrol et al. (2016)) in comparison to lattice data \cite{Boucaud:2018xup} (Boucaud et al. (2018)).
	The bands correspond to different solutions as explained in the text.}
	\label{fig:comp_gh_gl}
\end{figure}

In \fref{fig:comp_gh_gl}, we show the functional results from the FRG and the 3PI effective action in comparison 
to lattice results for the ghost and gluon dressing functions $G(p^2)$ and $Z(p^2)$, respectively. The functional 
results have been obtained in truncations on a similar level. Nevertheless, the corresponding equations are 
different and complementary in structure. Thus, the very good agreement between these methods and the agreement
with lattice gauge theory provides first evidence that this level of truncation is sufficient. 

\subsubsection{Three-gluon vertex}
\label{sec:tg}

The three-gluon vertex plays a central role for the calculation of glueballs as it constitutes the dominant interaction in the kernels, see Sec.~\ref{sec:approximations_bse}.
First functional calculations of the three-gluon vertex can be found in \cite{Alkofer:2008dt,Blum:2014gna,Eichmann:2014xya}.
Such calculations with fixed input have been subsequently refined, see, e.g., \cite{Aguilar:2021lke,Aguilar:2023qqd}.
Also STI-based calculations have been performed \cite{Aguilar:2019jsj}.
In \cite{Cyrol:2016tym,Huber:2020keu,Pawlowski:2022oyq}, the vertex was included as a dynamic quantity into a larger system of equations, leading to decent agreement with lattice results, as, for example, from \cite{Cucchieri:2008qm,Athenodorou:2016oyh,Duarte:2016ieu,Sternbeck:2017ntv,Boucaud:2017obn,Maas:2020zjp,Aguilar:2021lke,Pinto-Gomez:2022brg,Pinto-Gomez:2024mrk}.

\begin{figure}[h!]
	\includegraphics[width=0.45\textwidth]{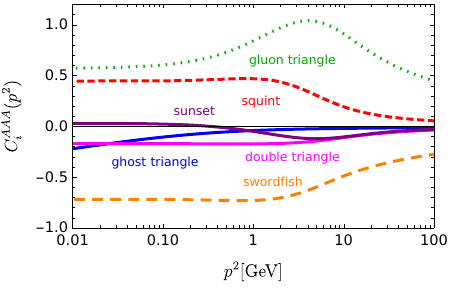}\hfill
	\caption{Contributions of diagrams of the three-gluon vertex DSE evaluated at the symmetric point.
	Topologically equivalent diagrams are summed up, e.g., the three swordfish diagrams.
	\label{fig:tgDiags}
	}

	\includegraphics[width=0.45\textwidth]{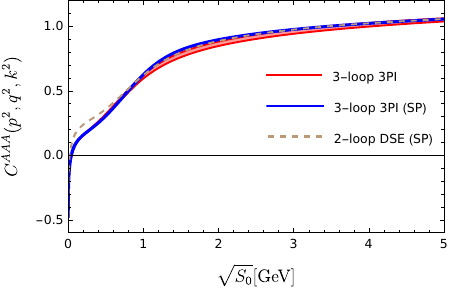}
    \caption{
    The three-gluon vertex dressing function $C^{AAA}(p^2,q^2,k^2)$ from the 3PI effective action truncated at three loops (solid lines) and its DSE including two-loop diagrams (dashed line) \cite{Huber:2020keu}.
    For 3PI, results calculated with full angular dependence and with dependence on $S_0$ evaluated at the symmetric point (SP) $S_0=p^2/2$ only are shown, the band indicating the angular dependence (variables $\rho$ and $\eta$ in \cite{Huber:2020keu}).
	}
    \label{fig:tg_DSE3PI_ang}
	
    \includegraphics[width=0.45\textwidth]{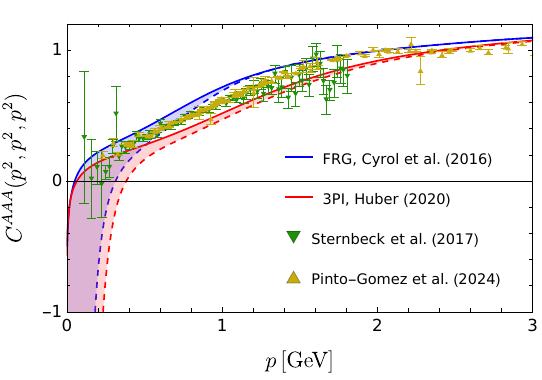}\\
    \caption{The three-gluon vertex at the symmetric point \cite{Huber:2020keu} in comparison to lattice data \cite{Sternbeck:2017ntv,Pinto-Gomez:2024mrk} and FRG results \cite{Cyrol:2016tym}.
	The bands correspond to the families of solutions.
    }
    \label{fig:tg_comp}
\end{figure}

Here we discuss the effect of truncations, compare results from different methods and discuss the role of kinematic approximations.
For the three-gluon vertex, it is known from DSE calculations that cancellations between diagrams appear.
In Fig.~\ref{fig:tgDiags} we show explicit contributions in its DSE including two-loop 
terms.\footnote{The relevance of cancellations is even more evident in two and three dimensions \cite{Huber:2012zj,Huber:2016tvc} because the dressings fall off polynomially instead of logarithmically in the UV.}
The ghost triangle leads to a zero crossing in the tree-level dressing function as it is negative and diverges in the IR \cite{Huber:2012zj,Pelaez:2013cpa,Aguilar:2013vaa,Blum:2014gna,Eichmann:2014xya,Huber:2016tvc}.
Since all diagrams contribute quantitatively, discarding any of it can have severe consequences.

An interesting finding is that the one-loop truncated DSE results (not shown) do not agree well with 3PI results.
In particular, the zero crossing of the vertex is at much higher momenta \cite{Eichmann:2014xya}.
This was remedied in earlier DSE calculations by effectively dressing the bare vertex with so-called renormalization group improvement factors \cite{Blum:2014gna}, the rationale being that this takes into account the perturbative resummation of higher-loop diagrams. However, this introduces some model dependence for the 
IR part.\footnote{Reassuringly, in two and three dimensions, where UV contributions drop power-like instead of logarithmically, the one-loop diagrams are sufficient to produce a reasonable zero crossing \cite{Huber:2012zj,Huber:2016tvc}, hinting at the important role of resummation in four dimensions.}
Indeed, when two-loop diagrams are included in the DSE (see Ref.~\cite{Huber:2018ned} for the diagrammatic representation), the zero crossing shifts to the IR and results become quantitative as illustrated in \fref{fig:tg_DSE3PI_ang}.
To alleviate the numeric effort, the two-loop DSE was solved in the symmetric point (SP) 
approximation where only a single variable is needed to describe the kinematics.
In the same plot, we show the consequence of this approximation for the 3PI calculation ('3PI' vs. '3PI (SP)').
Only in the midmomentum regime a small deviation is visible.
The reason for this approximation working so well is the weak angular dependence of the three-gluon vertex, shown explicitly in \fref{fig:tg_DSE3PI_ang}.
This property, also called planar degeneracy \cite{Pinto-Gomez:2022brg}, was already observed in \cite{Eichmann:2014xya} and subsequently thoroughly investigated \cite{Pinto-Gomez:2022brg,Pinto-Gomez:2024mrk}.
It is quite robust as we checked by plotting the results of the model dependent calculation of \cite{Blum:2014gna} with the appropriate kinematic 
variable\footnote{This is the definition from \cite{Eichmann:2014xya} where $S_0$ was introduced as singlet variable for the permutation group $S_3$. In the literature, sometimes also 1/3 instead of 1/6 is used which corresponds to the averaged squared momenta.}
\begin{align}
 S_0= \frac{p^2+q^2+k^2}{6},
\end{align}
where $p$, $q$ and $k$ are the momenta of the vertex legs.
Also for this simplified setup we found clear evidence of planar degeneracy.

After having established the good agreement between DSE and 3PI calculations, we compare now 
the 3PI result with FRG and lattice results in \fref{fig:tg_comp}.
For both functional methods, the colored bands correspond to the families of solutions discussed above.
Overall, the agreement between the three methods is qualitatively very good with sizeable quantitative
deviations only in the midmomentum regime. These need to be further explored in future work.

\subsubsection{Ghost-gluon vertex}
\begin{figure}[h!]
	\includegraphics[width=0.46\textwidth]{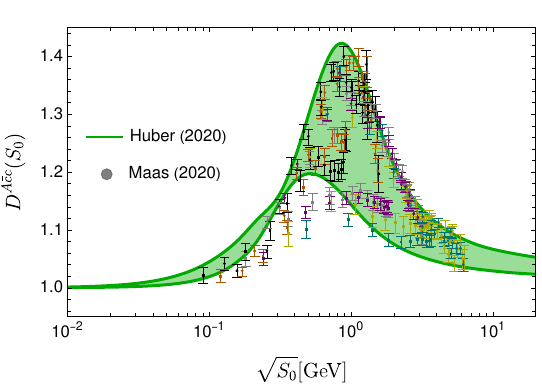}
    \caption{Ghost-gluon vertex dressing function $D^{A\bar c c}(k^2;p^2,q^2)$ as a function of $S_0 = (k^2+p^2+q^2)/6$  from functional and lattice methods \cite{Maas:2019ggf}. Here the bands correspond to the sizeable angular dependence of the dressing function.}
    \label{fig:ghg1} %
	\includegraphics[width=0.46\textwidth]{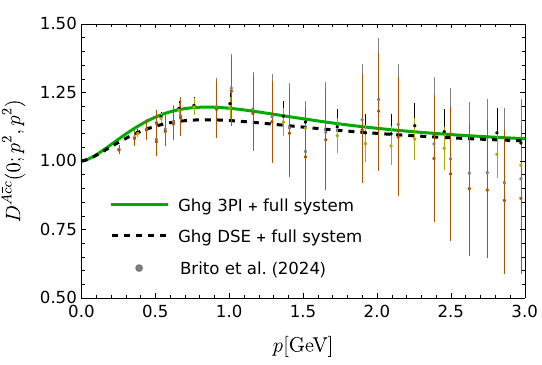}
	\caption{Ghost-gluon vertex dressing function $D^{A\bar c c}(0;p^2,q^2)$ in the soft gluon limit $k^2 \rightarrow 0$.
		Compared are lattice results \cite{Brito:2024aod} with results from functional methods once self-consistently from the 3PI effective 
		action and once using the 1PI DSE version of the EOM for the ghost-gluon vertex.
		\label{fig:ghg2} 
	} 
	\includegraphics[width=0.46\textwidth]{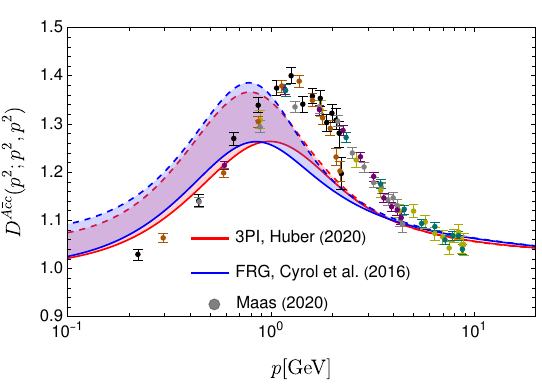}
	\caption{Ghost-gluon vertex dressing function $D^{A\bar c c}(p^2;p^2,p^2)$ at the symmetric point from 3PI \cite{Huber:2020keu}, FRG \cite{Cyrol:2016tym} and lattice calculations \cite{Maas:2019ggf}.
	The bands correspond to the families of solutions.
	\label{fig:ghg3} } 
\end{figure}

Motivated by Taylor's nonrenormalization theorem in Landau gauge \cite{Taylor:1971ff}, the ghost-gluon vertex 
has been assumed to be bare in many DSE studies. However, later explicit calculations showed the vertex dressing 
to deviate from one which are in turn relevant when the vertex appears in other equations; see, e.g.
\cite{Schleifenbaum:2004id,Dudal:2012zx,Huber:2012kd,Aguilar:2013xqa}, for early results.
Lacking the Bose symmetry of the three-gluon vertex, the ghost-gluon vertex shows a clear angular dependence 
as shown in \fref{fig:ghg1}. In \fref{fig:ghg2} we compare two different setups: the functional results displayed 
were obtained once from the equation of motion for the ghost-gluon vertex derived from the 3PI effective action 
and solved together with corresponding equations for all other correlation functions and once with the EOM for 
the ghost-gluon vertex replaced by the 1PI DSE variant. It is interesting to observe that the small but visible 
difference seen here does not propagate to the other correlation functions: these are remarkably stable and 
practically invariant \cite{Huber:2020keu}. This is again a signal for apparent convergence and the stability of
the truncation against perturbations. In \fref{fig:ghg3} we finally compare the 3PI results with corresponding 
FRG results and find remarkable agreement for the whole family of solutions given by the shaded bands. We also 
observe a sizeable shift in scale as compared to the lattice results of Ref.~\cite{Maas:2019ggf}.
This again needs to be further explored in future work.

\subsection{Extensions of truncation}
\label{sec:extensions}

In the previous sections we argued that functional methods, most notably the 3PI approach and the FRG calculations,
have reached a remarkable level of convergence with each other and with results from lattice gauge theory. 
Nevertheless, we also encountered quantitative discrepancies in specific kinematic regions that need to be explored
further. In the following we therefore discuss what we already know about the stability of the present 3PI 
truncation scheme against possible extensions. 

First, we need to discuss the structure of the two gluon self-interaction vertices. 
The three-gluon vertex has been studied in great detail both in functional methods and lattice gauge theory. 
It has been shown to be dominated by the tree-level tensor while the dressings of the other three tensors 
are heavily suppressed 
\cite{Eichmann:2014xya,Maas:2020zjp,Pinto-Gomez:2022brg,Pinto-Gomez:2024mrk}. Thus, for reasons of complexity 
and CPU time, their contributions to the full 3PI system have been neglected so far. 

\begin{figure}[tb]
	\includegraphics[width=0.48\textwidth]{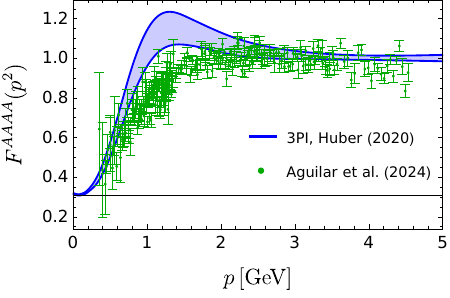}\hfill
	\caption{The four-gluon vertex \cite{Huber:2020keu} in the soft gluon limit in comparison with lattice data \cite{Aguilar:2024dlv}.
		The band corresponds to the remaining angular dependence.
		The functional data was renormalized at the symmetric point $p^2=4.3\,\text{GeV}^2$ and lattice data at the same value in soft kinematics, see \cite{Aguilar:2024dlv}, but the effect on the figure is negligible within the present error bars.
	}
	\label{fig:fg_comp}
\end{figure}

While this approximation is well justified for the three-gluon vertex, the situation may be different for 
the four-gluon vertex. Its structure is much more complex with a transverse 
basis of 41 (Lorentz) times 5 (color) tensors \cite{Pascual:1980yu,Eichmann:2015nra}. 
A comparison of the tree-level structure of the four-gluon vertex with lattice data is shown in \fref{fig:fg_comp} 
for the case of one vanishing gluon momentum. The remaining dependence on the absolute momentum scale $p$ is
plotted on the x-axis and the angular dependence is shown as a band. We observe an infrared suppression of this vertex
both for the functional results \cite{Huber:2020keu} and the ones from lattice gauge theory \cite{Aguilar:2024dlv}. 
In the midmomentum regime the functional results feature a bump, which is not present in the lattice data, but
overall, the agreement is very satisfactory.\footnote{
One should also note that the anomalous dimension of the four-gluon vertex is positive so that at higher momenta 
the dressing necessarily rises again as also observed in the functional results \cite{Huber:2020keu}.}

Studies of a selected few non-tree level dressing functions of the four-gluon vertex indeed indicate heavy suppression in the mid- and 
highmomentum regions
\cite{Kellermann:2008iw,Binosi:2014kka,Cyrol:2014kca,Huber:2016tvc,Aguilar:2024fen}. However, for intermediate momenta this does not need to be the case for all dressing functions.
In addition, some non-tree level structures feature interesting properties 
such as infrared divergences. Thus, for a definite answer on their overall importance, 
further studies are required. Fortunately, the impact of the four-gluon vertex on our 3PI truncation scheme 
is very limited. The only place this vertex appears is in the DSE for the gluon propagator. For this quantity,
however, we already observe excellent agreement between functional results and the ones from lattice gauge theory
even if only the tree-level dressing of the four-gluon vertex is taken into account \cite{Huber:2020keu}. We 
consider this as indirect evidence for either cancellations or indeed heavy suppression of all non-tree level 
structures in the four-gluon vertex.\footnote{ 
As an aside, we also note that the four-gluon vertex itself is of interest for glueballs as it contains the 
corresponding pole of a two-gluon bound state which can be extracted using appropriate spectral reconstruction 
techniques \cite{Pawlowski:2022zhh}.}

\begin{figure*}[tb]
	\begin{center}
		\includegraphics[width=0.48\textwidth]{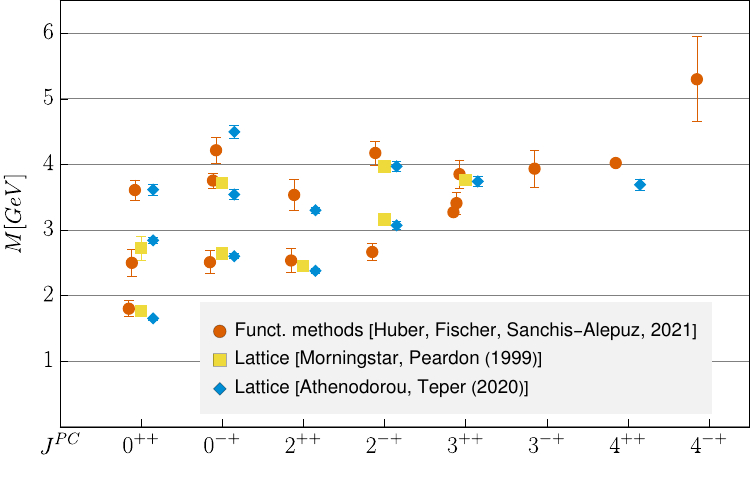}\hfill
		\includegraphics[width=0.48\textwidth]{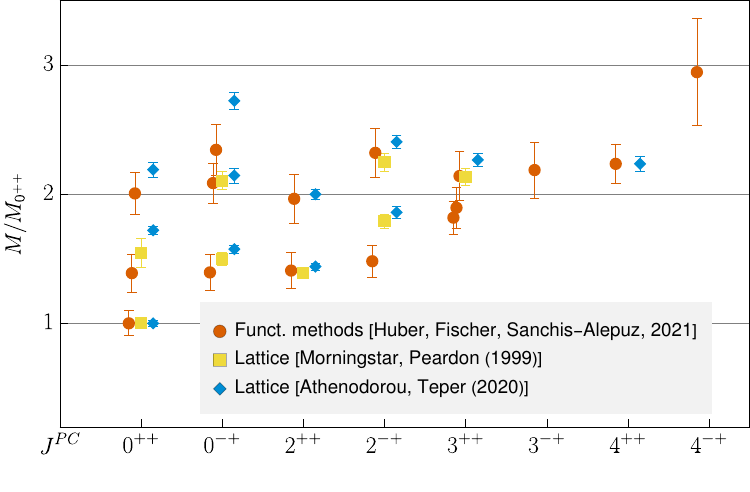}
	\end{center}
	\caption{
		Results for glueball ground states and excited states for the indicated quantum numbers from lattice simulations \cite{Morningstar:1999rf,Athenodorou:2020ani} and functional equations \cite{Huber:2021yfy,Huber:2022dsn}.
		Left: Absolute scale set by $r_0=1/(418(5)\,\text{MeV})$.
		Right: Spectrum relative to the ground state.
		See \cite{Huber:2021yfy} for more info on errors and the identification of states.
		The functional results have changed by a factor of $0.970$ compared to \cite{Huber:2021yfy} due to an improved scale setting for the input, see the appendix.
	}
	\label{fig:spectrum}
\end{figure*}

A qualitatively different extension is the inclusion of other four-point functions which we only mention here 
for completeness. Such functions, namely the two-ghost-two-gluon and four-ghost vertices, do not appear directly 
in the 3PI equations of motion, but their effects are encoded in higher loop terms.
Thus, it is easier to assess the effect of these correlation functions by studying their impact in 1PI DSEs where 
they can appear in one-loop diagrams \cite{Huber:2017txg}.
The aforementioned four-point functions have 25 (transverse) and 5 tensors, respectively.
Using a single kinematic configuration, all of these tensors were calculated in \cite{Huber:2017txg} and subsequently their effect was tested for the ghost-gluon, three-gluon and four-gluon vertices.
For the last two, the effect was practically negligible.
This can partially be traced to the fact that some tensors do not contribute due to their color structure.
Only for the ghost-gluon vertex a small effect of at most $2\,\%$ was found.
Hence, the inclusion of such four-point functions in DSEs, or equivalently the corresponding higher loop diagrams in equations of motion from the 3PI effective action, is not expected to be relevant at the current level of precision and within the employed kinematic approximation.
However, it should be noted that not only do such vertices appear differently in FRG equations due their inherent one-loop structure with only dressed vertices, but they also can have different effects there \cite{Corell:2018yil}.

\section{Glueball spectrum}
\label{sec:spectrum}

\begin{figure*}[tb]
	\includegraphics[width=0.48\textwidth]{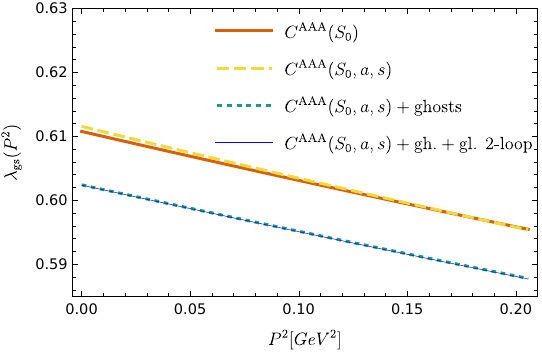}\hfill
	\includegraphics[width=0.48\textwidth]{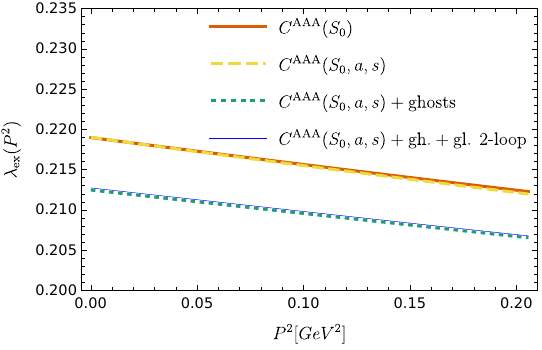}
	\caption{Eigenvalue curves for the ground state (left) and first excited state (right) of the scalar glueball for three approximations: only gluonic diagram with $S_0$ approximation for the three-gluon vertex (solid), only gluonic diagram with full angular dependence of the three-gluon vertex (dashed), all one-loop diagrams (dotted) and all one-loop diagrams plus gluonic two-loop diagrams (thin line).}
	\label{fig:eigenvalue_comp}
	\includegraphics[width=0.48\textwidth]{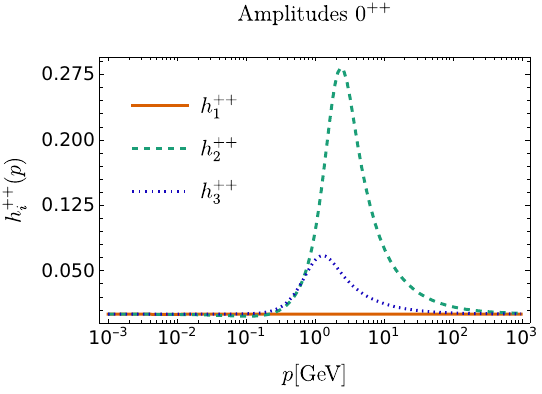}\hfill
	\includegraphics[width=0.48\textwidth]{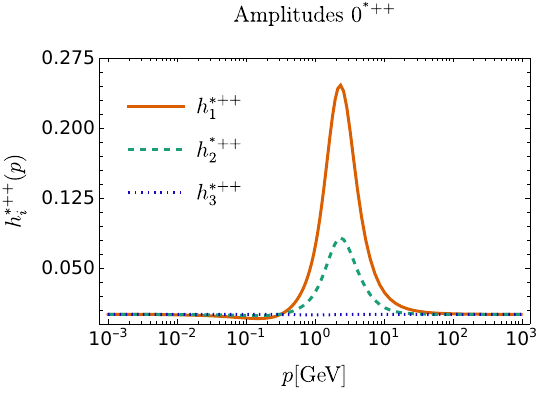}
	\caption{Leading amplitudes for the scalar ground (left) and first excited states (right). The amplitudes 
	 $h_{1,2}^{++}$ correspond to two gluon contributions, whereas $h_3^{++}$ is the amplitude of 
	 the ghost part.}
	\label{fig:amplitudes}
\end{figure*}

The very existence of pure glueball states is only possible due to the self-interaction of gluons. It is therefore highly
natural for the three-gluon vertex to play a crucial role for the glueball spectrum. This has been noted in model calculations
\cite{Meyers:2012ka,Sanchis-Alepuz:2015hma,Souza:2019ylx,Kaptari:2020qlt} and our previous results using a self-consistently
calculated vertex \cite{Huber:2020ngt,Huber:2021yfy}. 

As discussed in detail in Sec.~\ref{sec:correlationFunctions}, the underlying truncation is an expansion of the 3PI effective 
action on the three-loop level.
For systematic reasons, the derivation of the kernels of the glueball BSEs needs to be done on the same level.
They then contain (nonperturbative) one-particle exchange and one-loop diagrams, see \fref{fig:kernels}, leading to one- and two-loop diagrams in the BSEs, respectively.
The latter are much more complicated to calculate than the former which are typically the only diagrams considered for the quark sector of QCD

Thus, we focused initially on the one-loop diagrams in the BSEs \cite{Huber:2020ngt,Huber:2021yfy}.
The corresponding results are shown in \fref{fig:spectrum} and listed in \tref{tab:masses} in comparison with lattice results.
Overall, the two methods agree very well on a qualitative and quantitative level.
In particular, the usual picture of the scalar glueball being the lightest one and its first excitation and the pseudoscalar and tensor glueballs being roughly of the same mass is confirmed.

In the following, we discuss variations of this setup. We start with a study of possible further simplifications in Sec.~\ref{sec:approximations_bse}. In Sec.~\ref{sec:two-loop}, we explore the effect of including the two-loop diagrams. Finally, the role of gauge invariance within a class of Landau-type
gauges is discussed in Sec.~\ref{sec:family}.

The correlation functions necessary to solve the BSE, discussed in Sec.~\ref{sec:correlationFunctions},
are presently only available for space-like and real momenta. Extensions into the time-like momentum domain 
are computationally very expensive and so far only available for much simpler truncations schemes, see, e.g., 
\cite{Fischer:2020xnb} and references therein. As a consequence, we cannot solve the glueball BSEs at 
values of total momentum $P$ corresponding to the bound state masses, $M^2=-P^2$, as this would probe the correlation functions at complex arguments. We therefore resorted 
to a different strategy. The BSEs are treated as eigenvalue equations with an eigenvalue $\lambda(P^2)$. 
Since a bound state corresponds to $\lambda(P^2)=1$, it is sufficient to determine $\lambda(P^2)$ for many 
space-like $P^2>0$ and then extrapolate the resulting eigenvalue curve into the time-like momentum region. 
To this end we employ a Schlessinger continued fraction method \cite{Schlessinger:1968spm,Tripolt:2018xeo}. 
The quality of this procedure is discussed in detail in Ref.~\cite{Huber:2020ngt} where also a comparison 
with a direct calculation of a system solvable at time-like $P$ is shown. In the following, all error bands 
shown for the functional results stem from the extrapolation procedure.

\begin{table*}[tb]
	\begin{center}
		\begin{tabular}{|l||c|c|c|c|c|c|c|c|}
			\hline
			&  \multicolumn{2}{c|}{\cite{Morningstar:1999rf}} & \multicolumn{2}{c|}{\cite{Chen:2005mg}} & \multicolumn{2}{c|}{\cite{Athenodorou:2020ani}} & \multicolumn{2}{c|}{\cite{Huber:2021yfy}}\\   
			\hline
			State &  $M\, [\text{MeV}]$& $M/M_{0^{++}}$ & $M\, [\text{MeV}]$& $M/M_{0^{++}}$  &  $M\, [\text{MeV}]$& $M/M_{0^{++}}$ & $M\,[\text{MeV}]$ & $M/M_{0^{++}}$\\   
			\hline\hline
			$0^{++}$ & $1760 (50)$ & $1(0.04)$ & $1740(60)$ & $1(0.05)$ & $1651(23)$ & $1(0.02)$ & $1800 (120)$ & $1(0.1)$\\
			\hline
			$0^{^*++}$ & $2720 (180)^*$ & $1.54(0.11)^*$ & -- & -- & $2840(40)$ & $1.72(0.034)$ & $2500 (210)$ & $1.39(0.15)$\\
			\hline
			\multirow{2}{*}{$0^{^{**}++}$} & \multirow{2}{*}{--} & \multirow{2}{*}{--} & \multirow{2}{*}{--} & \multirow{2}{*}{--} & $3650(60)^\dagger$ & $2.21(0.05)^\dagger$ & \multirow{2}{*}{$3610 (150)$} & \multirow{2}{*}{$2.01(0.16)$}\\
			& & & & & $3580(150)^\dagger$ & $2.17(0.1)^\dagger$ & &\\
			\hline
			$0^{-+}$ & $2640 (40) $ & $1.50(0.05)$ & $2610(50)$ & $1.50(0.06)$ & $2600(40)$ & $1.574(0.032)$ & $2510 (170)$ & $1.39(0.14)$\\
			\hline
			$0^{^*-+}$ & $3710 (60)$ & $2.10(0.07)$ & -- & -- & $3540(80)$ & $2.14(0.06)$ & $3750 (110)$ & $2.09(0.16)$\\
			\hline
			\multirow{2}{*}{$0^{^{**}-+}$} & \multirow{2}{*}{--} & \multirow{2}{*}{--} & \multirow{2}{*}{--} & \multirow{2}{*}{--} & $4450(140)^\dagger$ & $2.7(0.09)^\dagger$ & \multirow{2}{*}{$4210 (200)$} & \multirow{2}{*}{$2.34(0.19)$}\\
			& & & & & $4540(120)^\dagger$ & $2.75(0.08)^\dagger$ & &\\
			\hline
			\hline
			$2^{++}$ & 2447(25) & 1.39(0.04) & 2440(50) & 1.40(0.06) & 2376(32) & 1.439(0.028) & 2530(180) & 1.41(0.14)\\
			\hline
			$2^{^*++}$ & -- & -- & -- & -- & 3300(50) & 2(0.04) & 3530(230) & 1.96(0.19)\\
			\hline
			$2^{-+}$ & 3160(31) & 1.79(0.05) & 3100(60) & 1.78(0.07) & 3070(60) & 1.86(0.04) & 2660(130) & 1.48(0.13)\\
			\hline
			$2^{^*-+}$ &  3970(40)$^*$ & 2.25(0.07)$^*$ & -- & -- & 3970(70) & 2.4(0.05) & 4170(180) & 2.32(0.19)\\
			\hline\hline
			$3^{++}$ & 3760(40) & 2.13(0.07) & 3740(60) & 2.15(0.09) & 3740(70)$^*$ & 2.27(0.05)$^*$ & 3270(50)$^*$ & 1.82(0.13)$^*$\\
			\hline
			$3^{^*++}$ & -- & -- & -- & -- & -- & -- & 3410(170)$^*$ & 1.89(0.16)$^*$\\
			\hline
			$3^{^{**}++}$ & -- & -- & -- & -- & -- && 3850(220)$^*$ & 2.14(0.19)$^*$\\
			\hline
			$3^{-+}$ & -- & -- & -- & -- & -- & -- & 3930(280)$^*$ & 2.19(0.22)$^*$\\
			\hline
			\hline
			$4^{++}$ & -- & -- & -- & -- & 3690(80)$^*$ & 2.24(0.06)$^*$ & 4020(20)$^*$ & 2.23(0.15)$^*$\\
			\hline
			$4^{-+}$ & -- & -- & -- & -- & -- & -- & 5300(600)$^*$ & 2.9(0.4)$^*$\\
			\hline
		\end{tabular}
		\caption{Ground and excited state masses $M$ of glueballs for various quantum numbers with the updated scale in the functional calculations.
			Compared are lattice results from \cite{Morningstar:1999rf,Chen:2005mg,Athenodorou:2020ani} with the results of \cite{Huber:2021yfy} including the update from \cite{Huber:2022dsn}.
			For \cite{Morningstar:1999rf,Chen:2005mg}, the errors are the combined errors from statistics and the use of anisotropic lattices.
			For \cite{Athenodorou:2020ani}, the error is statistical only.
			In our results, the error comes from the extrapolation method and should be considered a lower bound on errors.
			All results use the same value for $r_0=1/(418(5)\,\text{MeV})$.
			The functional results have changed by a factor of $0.970$ compared to \cite{Huber:2021yfy} due to an improved scale setting for the input, see the appendix.
			The related error is not included in the table.
			Masses with $^\dagger$ are conjectured to be the second excited states.
			Masses with $^*$ come with some uncertainty in their identification in the lattice case or in the trustworthiness of the extrapolated value in the BSE case.
		}
		\label{tab:masses}
	\end{center}
\end{table*}

\subsection{Approximations within one-loop truncation}
\label{sec:approximations_bse}

As stated above, the glueball spectrum of Ref.~\cite{Huber:2020ngt,Huber:2021yfy} was obtained using the full
3PI results for the correlation functions as input, but using a one-loop truncation of the BSE kernels displayed in
Fig.~\fref{fig:kernels} (yellow boxes). Let us first discuss, whether this one-loop truncation could be simplified.  
All results of this subsection are discussed for the example of the ground state and first excited state of the 
scalar glueball only. 
 
We start with the simplest possible setup, neglecting all ghost effects and the four-gluon vertex in 
the gluon part of the glueball BSEs. The resulting BSE only contains the one-gluon exchange diagram and
resembles in structure corresponding BSEs in the quark sector. In addition, we test the relevance of the 
kinematic angular dependence of the three-gluon vertex. As discussed in Sec.~\ref{sec:tg}, this angular 
dependence, encoded in the variables $a$ and $s$ defined in \cite{Eichmann:2014xya}, is very weak and may be neglected by treating the three-gluon vertex as a function of only 
one variable for which we choose here $S_0=(p^2+q^2+k^2)/6$. The resulting eigenvalue curves at 
space-like momenta are shown in \fref{fig:eigenvalue_comp}. Indeed, we find that the two eigenvalue curves
without ghosts, $C^{AAA}(S_0)$ without angular dependence and $C^{AAA}(S_0,a,s)$ including the dependence
on the angular variables $a,s$, are almost indistinguishable. However, since the small deviations are momentum 
dependent, this entails a mass difference of the order of $50\,\text{MeV}$ in the scalar glueball mass after 
extrapolation. It therefore depends on the desired precision goal whether the angular dependence in the 
three-gluon vertex can be neglected or not. This is different for the four-gluon vertex; its addition to the
kernel has much smaller effects on the sub-permille level.

The largest effect, however, stems from the ghost diagrams in the kernel. The green dashed curves in 
\fref{fig:eigenvalue_comp} ($C^{AAA}(S_0,a,s)$ + ghosts) represent results from the full one-loop 
truncation discussed above including 
all ghost terms. We observe a sizeable shift in the eigenvalue curves at space-like momenta, which results 
in extrapolated masses at time-like momenta differing by about $200\,$MeV for the scalar glueball ground state 
and by about $350\,$MeV for the excited state. This makes it mandatory to include
the ghost diagrams. Interestingly, the glueball amplitudes displayed in \fref{fig:amplitudes} show a clear 
hierarchy which, however, is different for ground and excited states. In particular, from the excited state 
one could think that the ghosts are not relevant because the amplitude is severely suppressed compared to 
the other ones. The results for the masses, however, show that this is not the case. Another interesting 
finding is that the extrapolation is more stable when ghosts are included and the associated error goes down
considerably.

To summarize: These results show that reducing the kinematic dependence of the three-gluon vertex to a single 
variable is a useful approximation that leads to a speed-up in the calculations while producing still quantitative 
results on a few percent level. Neglecting the ghost contributions, on the other hand, is only qualitatively 
acceptable and leads to a noticeable shift in the glueball masses.

\begin{figure*}[t!]
	\includegraphics[width=0.48\textwidth]{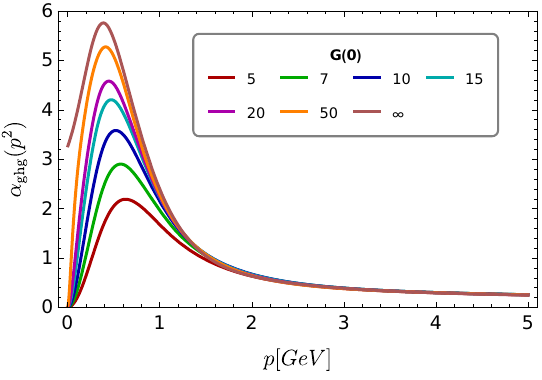}\hfill
	\includegraphics[width=0.48\textwidth]{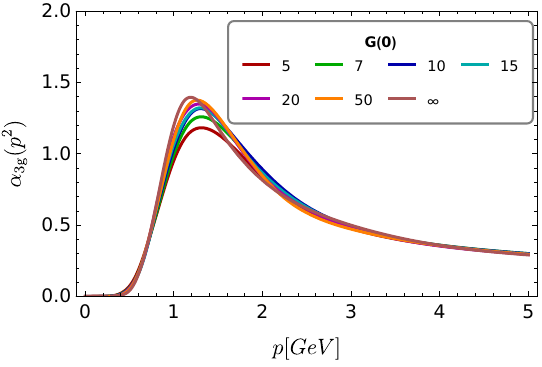}
	\caption{Couplings from the ghost-gluon (left) and three-gluon (right) vertices for different solutions labelled by the IR value of the ghost dressing function $G(0)$.}
	\label{fig:couplingsFamily}
\end{figure*}

\subsection{Two-loop truncation}\label{sec:two-loop}

As discussed above, full consistency of the BSEs with our 3PI system for the correlation functions is only achieved 
including the two-loop diagrams in the kernel, \fref{fig:kernels}. Since these are considerably more complicated
to calculate, we resort for now to the following simplifications:
First, we only use the three-gluon vertex for symmetric kinematics in the two-loop diagrams, neglecting its
angular dependence. As discussed above, this is a very good approximation and helps to speed up the calculations 
considerably. Second, we use a numerical setup with lower precision. We tested the impact of this setup for the 
one-loop case and found that it allows for a reliable extraction of the ground and first excited states. 
Higher excited states become unstable and are consequently not considered. Third, given the dominant role of the 
purely gluonic diagram, we added only the two-loop diagrams in the gluon-gluon scattering kernel, i.e. the black
box in \fref{fig:kernels}.

We performed the calculations for the scalar, pseudo\-scalar and tensor glueballs with first results reported previously 
in conference proceedings \cite{Huber:2022rhh,Huber:2021zqk,Huber:2023mls}.
The resulting eigenvalue curves have also been included in \fref{fig:eigenvalue_comp} 
($C^{AAA}(S_0,a,s)$ + gh. + gl. 2-loop) and are hardly distinguishable from the ones of the one-loop calculations. 
Also the extrapolations showed no relevant change. 
The largest effect was seen for the scalar glueball mass and of the order of 2\,\% 
and hence well below the general error of the extrapolation procedure.

Hence, the existing results remain valid and we conclude that the sum of the gluonic two-loop diagrams does not 
play a significant role. In order to decide whether cancellations between the two-loop diagrams occur, we repeated 
the calculations also with only single two-loop diagrams instead of all of them. Again, the results did not change
appreciably, and we therefore conclude that also individual gluonic two-loop diagrams have only tiny effects.
This is in marked contrast to the three-gluon vertex equation discussed in Sec.~\ref{sec:tg} where sizeable 
cancellations between different diagrams occur.

\subsection{Family of solutions and gauge invariance}
\label{sec:family}

As explained in Sec.~\ref{sec:correlationFunctions}, the DSEs of Yang-Mills theory are solved by a 
one-parameter family of correlation functions. In the literature, this family has been interpreted as a manifestation
of a residual gauge freedom arising from incomplete gauge fixing in Landau gauge due to Gribov ambiguities
\cite{Fischer:2008uz,Maas:2009se}. Other authors have hypothesized that different members of this family are not 
gauge-equivalent and one has to identify the 'correct' one by a physical condition, e.g., by minimizing the 
effective potential \cite{Dudal:2011gd}. Since the spectrum of Yang-Mills theory must be gauge independent 
one can use it as a testing ground for both ideas: in the former case one would expect the same spectrum (within 
uncertainties) for the complete family, whereas in the second case the results should show significant deviations.

We solved the BSEs for seven different cases to be distinguished here by the value of the ghost dressing function at vanishing momentum, $G(0)$.
The corresponding couplings of the ghost-gluon and three-gluon vertices \cite{Alles:1996ka,Alkofer:2004it,Eichmann:2014xya},
\begin{subequations}
	\label{eq:couplings}
	\begin{align}
	 \alpha_\text{ghg}(p^2)&=\alpha(\mu^2)\left[D^{A\bar cc}(p^2)\right]^2[G(p^2)]^2Z(p^2),\\
	 \alpha_\text{3g}(p^2)&=\alpha(\mu^2)\left[C^{AAA}(p^2)\right]^2[Z(p^2)]^3,
	\end{align}
\end{subequations}
respectively, are shown in \fref{fig:couplingsFamily}. They agree well in the perturbative regime and start to deviate around $2\,\text{GeV}$, reflecting a general observation that ambiguities due to gauge fixing are always related to
the infrared behavior of correlation functions. 
Since these couplings provide direct information about the interaction strength in the kernels of the BSE, it is
a very interesting and nontrivial question whether different solutions from this family also lead to different
glueball masses.

An important technical point of our calculations regard the proper numerical handling of the three-gluon vertex.
This vertex develops an infrared divergence which changes its nature within the family: with increasing $G(0)$ 
it becomes stronger and switches from logarithmic \cite{Aguilar:2013vaa,Cyrol:2016tym,Huber:2020keu} to power-like 
for the scaling limit $G(0) \rightarrow \infty$ \cite{Alkofer:2004it,Fischer:2009tn,Fischer:2006vf,Fischer:2008uz,Blum:2014gna,Eichmann:2014xya,Cyrol:2016tym,Huber:2020keu}.
This requires an increase in numerical precision as well as a specialized interpolation routine with the vertex dressing 
mapped to an $arcsinh$ function to handle the divergence properly.

Another technical point concerns the extrapolation procedure and the associated physical scale. As with all calculations
in Yang-Mills theory all results are first obtained in internal units. Physical units are set only at the end by the peak 
of the gluon dressing function, see \ref{sec:scale} for details. Since the corresponding scale factors are different for
different members of the family, the intervals for the extrapolation in $P^2$ also differ. To ensure best comparability, 
we only used values for $P^2$ within an interval common to all calculations and the same numbers of points for 
all extrapolations. This reduces the number of points available for the extrapolation to some extent, which explains 
the small deviations from and partially larger errors than in the high precision results shown in \fref{fig:spectrum}.

\begin{figure}[tb]
	\includegraphics[width=0.48\textwidth]{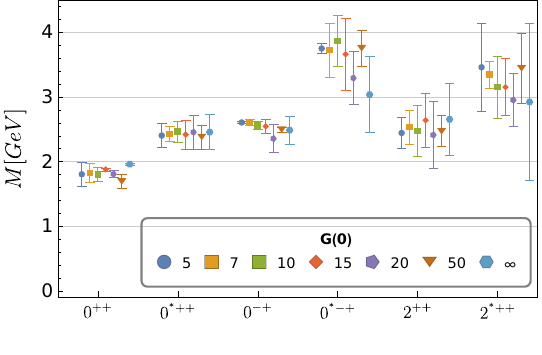}
	\caption{Masses for scalar, pseudoscalar and tensor ground and excited states from different input labelled by the IR value of the ghost dressing function $G(0)$.}
   \label{fig:spectrum_family}
\end{figure}

Scrutinizing our results, it is interesting to note that the eigenvalue curves obtained from all members of the family 
are qualitatively similar but differ quantitatively by a few percent. These are, however, not gauge invariant and
therefore only the extrapolated masses provide a meaningful estimate of the physical equivalence of the different 
solutions. The resulting spectrum of scalar, pseudoscalar and tensor glueballs is shown in \fref{fig:spectrum_family}. 
Given that the underlying correlation functions of the family differ drastically, cf. the shaded bands discussed in 
Sec.~\ref{sec:correlationFunctions}, it is striking and highly nontrivial how well the glueball masses align across 
the family. Within the extrapolation errors, we do not observe any significant variations. This agreement strongly
favors the proposition that different solutions within the family are physically equivalent. While this agreement 
might in principle be a 'coincidence', we do not believe this to be the case. To test this further, we also calculated 
the glueball masses with 'mixed' input, viz., we used propagators and vertices from different members of the family of
solutions. For solutions far enough apart, the mass of the scalar glueball dropped significantly, e.g., for mixing 
solutions $G(0)=5$ and $G(0)=20$ the ground state mass is below $1\,\text{GeV}$. For the $J=0$ glueballs, this investigation of the gauge dependence
was already done previously \cite{Huber:2020ngt}. Here we show the quantitative comparison and extend it to the 
tensor glueball.

These results strongly support the hypothesis of the different solutions corresponding to different gauges \cite{Maas:2009se}.
For a final conclusion, however, more insight into the nonperturbative gauge fixing ambiguity is needed, and more physical 
observables should be calculated.

\section{Summary and discussion}
\label{sec:summary}

In the past twenty years, functional methods have seen enormous progress. Theoretical tools are available that range
from simple model applications to high quality approximations of the full theory using systematic expansions. In this
work, we discussed in detail one of the most systematic approximation schemes applied so far based on the 3PI effective
action in (nonperturbative) three-loop expansion. We discussed the stability of this system under various modifications
and extensions in Sec.~\ref{sec:correlationFunctions}. Except for an unproblematic kinematic approximation of the 
three-gluon vertex based on planar degeneracy any further simplification of this setup leads to unwanted qualitative 
and quantitative changes. In contrast, any extensions tested so far only entail minor quantitative ones. For the current 
precision goal of our functional calculations on the five percent level the employed setup thus constitutes a very 
reliable basis. We consider the good agreement between different functional methods (3PI, 1PI DSE and FRG) and lattice
gauge theory as well as the observed stability under modifications as an indication of apparent convergence of the 
functional system of equations. Taking into account all primitively divergent correlation functions seems to be 
a necessary and sufficient condition for a reliable truncation. 

This high quality input also seems to be necessary and sufficient for a high quality determination of the glueball
spectrum \cite{Huber:2020ngt,Huber:2021yfy}, provided the kernels in the glueball BSEs match this complexity. 
We discussed the truncation of the BSE in Sec.~\ref{sec:spectrum} and explored the hierarchy of gluonic vs. ghost 
diagrams as well as the impact of hitherto neglected two-loop diagrams in the gluon part of the BSE. 
We found that the ghost diagrams do have a relevant quantitative impact, while effects of the two-loop diagrams 
are remarkably small. To our mind, this is another indication of apparent convergence.

Finally, we quantified the effect of using different members of the family of solutions accessible with functional 
methods. Within the extrapolation errors, the results agree, supporting the conjecture that different solutions 
correspond to different nonperturbative completions of Landau type gauges.

The results for the glueball spectrum and the tests performed here motivate investigations beyond this setup.
One possibility is the extension to three-gluon glueballs which would provide access to additional quantum numbers.
Another, the physically most relevant one, is the inclusion of quarks to investigate the mixing between quark-antiquark 
mesons and glueballs. It would certainly be most interesting to investigate this in the light of the recent analyses 
of experimental data \cite{Sarantsev:2021ein,Rodas:2021tyb,Klempt:2022qjf,BESIII:2023wfi}. This is material for future work.

\section*{Acknowledgments}
We are grateful to Feliciano de Soto, Mauricio Ferreira, and Orlando Oliveira for discussions and/or sharing of their data.
This work was supported by the DFG (German Research Foundation) grant FI 970/11-2. This work has also been supported by 
Silicon Austria Labs (SAL), owned by the Republic of Austria, the Styrian Business Promotion Agency (SFG), the federal 
state of Carinthia, the Upper Austrian Research (UAR), and the Austrian Association for the Electric and Electronics 
Industry (FEEI).

\appendix

\section{Scale setting}
\label{sec:scale}

In many functional calculations, a physical scale enters explicitly by using a model or input from other sources.
Here, however, the correlation functions are calculated in a self-contained way with no such explicit scale.
Hence, the raw results are in internal units and have to be converted to 'physical' ones using external input.
In Yang-Mills theory, glueball masses can serve as physical scale, as they are gauge invariant.
This is used in one of the spectrum plots in \fref{fig:spectrum} where the scale is set by the scalar glueball mass.
It is more convenient, though, to express the results in terms of GeV.
To this end, we take over the scale from lattice calculations by matching a distinct quantity between lattice and functional results.
Due to the good agreement of the gluon propagator, we use the peak of the gluon dressing function for this purpose.
In \cite{Huber:2020keu}, we used the lattice results of \cite{Sternbeck:2006rd} with $r_{0,\text{S}}=0.5\,\text{fm}=2.534/\text{GeV}$ and the position of the maximum at $p_\text{max,S}^2=0.94\,\text{GeV}^2$.
However, different lattice calculations may use different values of $r_0$.
This has to be taken into account here, because the lattice results of \cite{Athenodorou:2020ani} use $r_{0,\text{AT}}=0.472\,\text{fm}=2.392/\text{GeV}$.
This leads to a factor of $2.534/2.392=1.059$ for translating the scale of the BSE calculations to the scale of the lattice results.
Also the lattice results of \cite{Morningstar:1999rf}, with $r_{0,\text{MP}}=0.480\,\text{fm}=2.439/\text{GeV}$, were translated to this scale.

Since newer lattice results for the gluon propagator have become available, we compared the scale inferred from them using the same technique.
For the results from \cite{Boucaud:2018xup}, we found that the peak position can be determined more accurately as $p_\text{max,B}^2=0.83\,\text{GeV}^2$.
All results for the correlation functions presented here use this updated scale setting.
For the glueball masses, we again translated the scale of the input, based on $r_{0,\text{B}}=0.515\,\text{fm}=2.616/\text{GeV}$ \cite{Boucaud:2018xup}, to $r_{0,\text{AT}}$ of \cite{Athenodorou:2020ani} by rescaling the original results for the masses in \cite{Huber:2020ngt,Huber:2021yfy} with
\begin{align}
r_{\text{update}}=\sqrt{\frac{0.83}{0.94}}\left(\frac{r_{0,\text{S}}}{r_{0,\text{AT}}}\right)^{-1}\frac{r_{0,\text{B}}}{r_{0,\text{AT}}}=0.970.
\end{align}
All quoted results for glueball masses use this scale setting with $r_0=r_{0,\text{AT}}=0.472\,\text{fm}=2.392/\text{GeV}$.

\bibliographystyle{utphys_mod}
\bibliography{literature_apparentConvergence}

\end{document}